\documentclass[8.5pt,twoside,twocolumn]{article}
\usepackage[super,sort&compress,comma]{natbib} 
\usepackage{mhchem}
 \usepackage{times}
\usepackage{sectsty}
\usepackage{balance} 
\usepackage{graphicx} 
\usepackage{lastpage}
\usepackage[format=plain,justification=raggedright,singlelinecheck=false,font=small,labelfont=bf,labelsep=space]{caption} 
\usepackage{fancyhdr}
\usepackage{amssymb}

\begin{document}

\thispagestyle{plain}
\fancypagestyle{plain}{
\fancyhead[L]{\includegraphics[height=8pt]{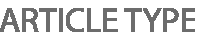}}
\fancyhead[C]{\hspace{-1cm}\includegraphics[height=20pt]{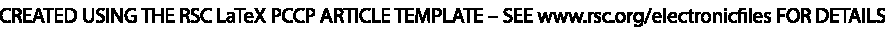}}
\fancyhead[R]{\includegraphics[height=10pt]{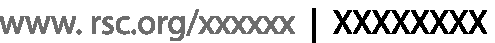}\vspace{-0.2cm}}
\renewcommand{\headrulewidth}{1pt}}
\renewcommand{\thefootnote}{\fnsymbol{footnote}}
\renewcommand\footnoterule{\vspace*{1pt}%
\hrule width 3.4in height 0.4pt \vspace*{5pt}} 
\setcounter{secnumdepth}{5}

\makeatletter 
\def\subsubsection{\@startsection{subsubsection}{3}{10pt}{-1.25ex plus -1ex minus -.1ex}{0ex plus 0ex}{\normalsize\bf}} 
\def\paragraph{\@startsection{paragraph}{4}{10pt}{-1.25ex plus -1ex minus -.1ex}{0ex plus 0ex}{\normalsize\textit}} 
\renewcommand\@biblabel[1]{#1}            
\renewcommand\@makefntext[1]%
{\noindent\makebox[0pt][r]{\@thefnmark\,}#1}
\makeatother 
\renewcommand{\figurename}{\small{Fig.}~}
\sectionfont{\large}
\subsectionfont{\normalsize} 

\fancyfoot{}
\fancyfoot[LO,RE]{\vspace{-7pt}\includegraphics[height=9pt]{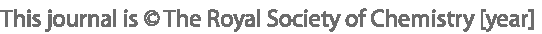}}
\fancyfoot[CO]{\vspace{-7.2pt}\hspace{12.2cm}\includegraphics{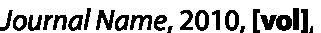}}
\fancyfoot[CE]{\vspace{-7.5pt}\hspace{-13.5cm}\includegraphics{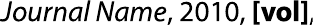}}
\fancyfoot[RO]{\footnotesize{\sffamily{1--\pageref{LastPage} ~\textbar  \hspace{2pt}\thepage}}}
\fancyfoot[LE]{\footnotesize{\sffamily{\thepage~\textbar\hspace{3.45cm} 1--\pageref{LastPage}}}}
\fancyhead{}
\renewcommand{\headrulewidth}{1pt} 
\renewcommand{\footrulewidth}{1pt}
\setlength{\arrayrulewidth}{1pt}
\setlength{\columnsep}{6.5mm}
\setlength\bibsep{1pt}

\twocolumn[
  \begin{@twocolumnfalse}
\noindent\LARGE{\textbf{Optically-isotropic responses induced by discrete rotational symmetry of nanoparticle clusters}}
\vspace{0.6cm}

\noindent\large{\textbf{Ben Hopkins,* Wei Liu, Andrey E. Miroshnichenko and Yuri S. Kivshar}}\vspace{0.5cm}

\noindent\textit{\small{\textbf{Received Xth XXXXXXXXXX 20XX, Accepted Xth XXXXXXXXX 20XX\newline
First published on the web Xth XXXXXXXXXX 200X}}}

\noindent \textbf{\small{DOI: 10.1039/b000000x}}
\vspace{0.6cm}

\noindent \normalsize{Fostered by the recent progress of the fields of plasmonics and metamaterials, the seminal topic of light scattering by clusters of nanoparticles is attracting enormous renewed interest gaining more attention than ever before.  Related studies have not only found various new applications in different branches of physics and chemistry, but also spread rapidly into other fields such as biology and medicine. Despite the significant achievements, there still exists unsolved but vitally important challenges of how to obtain robust polarisation-invariant responses of different types of scattering systems.  
In this paper, we demonstrate polarisation-independent responses of any scattering system with a rotational symmetry with respect to an axis parallel to the propagation direction of the incident wave. We demonstrate that the optical responses such as extinction, scattering, and absorption, can be made independent of the polarisation of the incident wave for all wavelengths. Such polarisation-independent responses are proven to be a robust and generic feature that is purely due to the rotational symmetry of the whole structure. We anticipate our finding will play a significant role in various applications involving light scattering such as sensing, nanoantennas, optical switches, and photovoltaic devices.}
\vspace{0.5cm}
 \end{@twocolumnfalse}
  ]

\footnotetext{\textit{Nonlinear Physics Center, Research School of Physics and Engineering,
Australian National University, Canberra ACT 0200, Australia.  \\E-mail: bth124@physics.anu.edu.au}}

\section{Introduction}

The current surging interest in various applications of nanoscale light-matter interactions, including biosensing~\cite{Jun2009_PNAS,Liu2011_Science,Kabashin2009_NM}, nanoantennas~\cite{Novotny2011_NP, Curto2010_Science}, photovoltaic devices~\cite{Atwater2010_NM} and many others, has triggered enormous effort into the old and fundamental problem of the manipulation of a particle's scattering and absorption characteristics~\cite{Bohren1983_book,Kreigib1995_book}. In the recently emerging fields of nanophotonics, various novel phenomena have been demonstrated involving interaction of nanoparticles with light, such as super-scattering~\cite{Ruan2010_PRL,Verslegers2012_PRL}, control of the direction of the scattered light by metasurfaces~\cite{Ni2011_science,Yu2011_science}, coherent perfect absorption of light by surface plasmons~\cite{Noh2012_PRL}, Fano resonances in nanoscale structures~\cite{Miroshnichenko2010_RMP,Lukyanchuk2011_NM} and plasmonic oligomers\cite{Brandl2006_JPCB,Hao2008_NL,Hentsche2010_NL,Hentsche2011_acsnano,King2011_ACSNANO,Ye2012_NL,Frimmer:PRL:2012}. At the same time, the interest in artificial magnetic responses that was fostered by the field of metamaterials has lead to the observation of artificial magnetic modes in nanoparticles and, since then, many related novel scattering features based on the interplay of both electric and magnetic responses have been demonstrated~\cite{Evlykuhin2010_PRB,Garixia_etxarri2011_OE,Kuznetsov2012_SciRep,Evlyukhin2012_NL,Miroshnichenko2012_ACSNANO,Liu2012_ACSNANO,Liu2012_PRB}.

To make further breakthroughs in different applications based on particle scattering, there is a fundamental challenge to overcome: polarisation dependence. The dependence of an optical response on polarisation comes from the fact that most structures have dominantly electric responses, which are highly dependent on the polarisation of the incident field. The simplest structure that does not exhibit polarisation-dependent scattering properties is a single spherical particle. According to the Mie theory the total extinction, scattering and absorption cross-sections do not depend on the incident polarisation angle, although the scattering diagram will exhibit some angle-dependent properties~\cite{Bohren1983_book}. It is possible to achieve a polarisation-independent scattering diagram by overlapping the electric and magnetic dipole responses of a single spherical nanoparticle~\cite{Liu2012_ACSNANO,Liu2012_PRB}, but such effects can only be achieved by rigorous engineering of the structure and can only happen in specific spectral regimes. However, it has also been experimentally observed that some plasmonic oligomer structures with discrete symmetries exhibit completely polarisation-independent extinction cross-section spectrums.~\cite{Hentsche2011_acsnano, Rahmani:N:2011, Chuntonov2011_NL}. This all leads to the question of what the necessary conditions are for an arbitrary system to have polarisation-independent scattering properties.

Inspired by the concepts of symmetry-induced degenerate states in quantum mechanics~\cite{Shankar1980_book} and mode degeneracy in uniform waveguides~\cite{Mcisaac1975_IEEE,Steel2001_OL,Liu2010_APL}, there have been some studies about symmetry-induced polarisation-independent scattering by clusters of particles~\cite{Brandl2006_JPCB,Aydin2011_NC,Sheikholeslmi2011_NL,Yurkin2010,Alsawafta2012}. However, as far as we know, there are no rigorous and systematic investigations of this topic. Additionally, in previous studies, usually only the dependence of extinction or scattering spectra on polarisation is investigated and the intrinsic loss spectra is neglected, which can be quite important in its own right, particularly for photovoltaic devices and biological applications.

\begin{figure}[t]
\centering
\centerline{\includegraphics[width=\columnwidth]{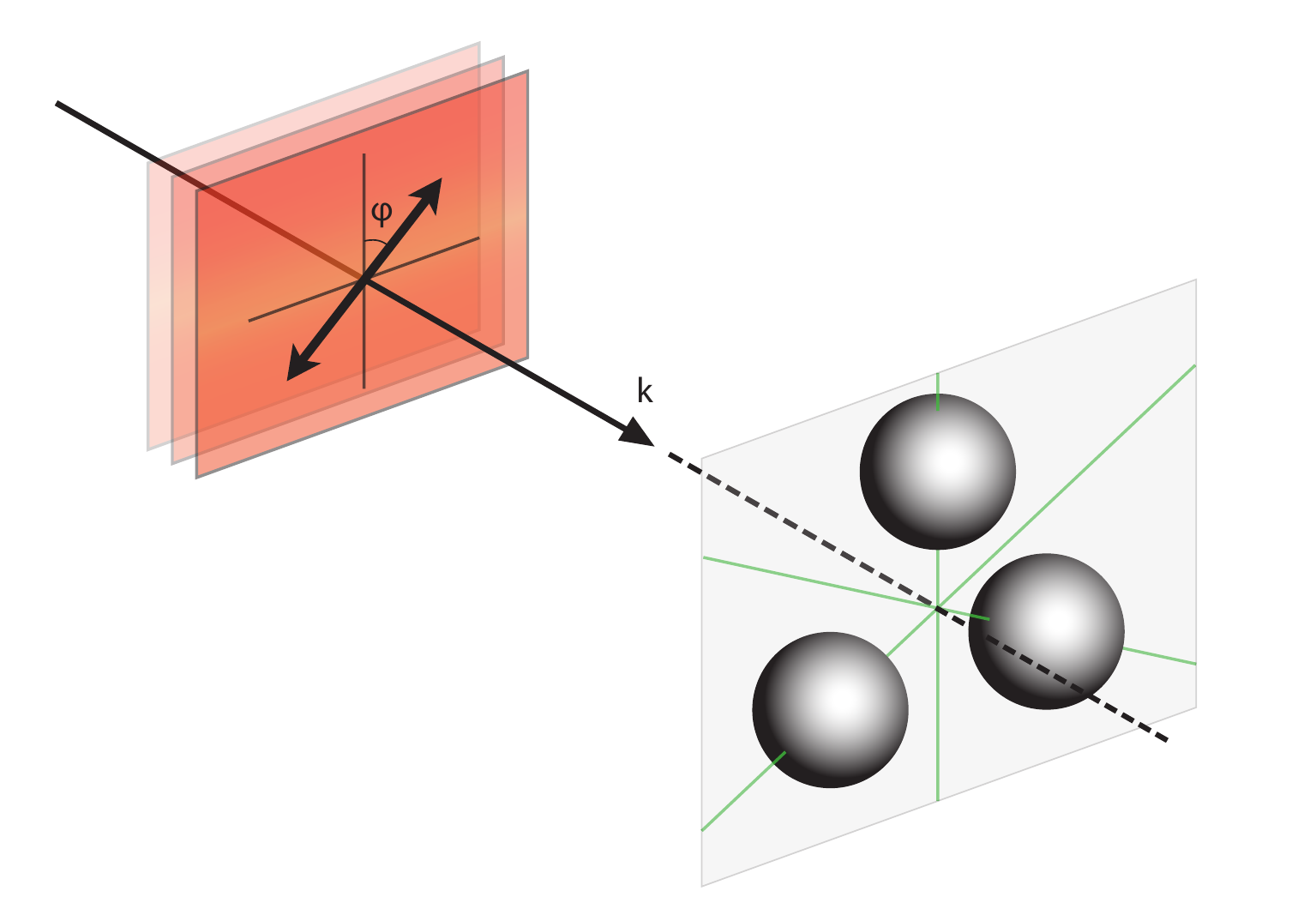}}
\caption{Schematic view of the problem under consideration: linearly-polarised plane wave scattering by a system with rotational symmetry.}
\label{fig:fig1}
\end{figure}

In this paper we show that established group theory concepts\cite{McWeeny1963,Cotton1990} can be used to deduce the effect of symmetry on discrete dipole scattering systems.  We then perform a systematic investigation on the optical responses of structures with an  {\em n}-fold ($n\geq3$) rotational symmetry, where the {\em n}-fold symmetry axis is parallel to the direction of propagation of the incident plane wave [see Fig.~\ref{fig:fig1}]. Such  {\em n}-fold symmetry implies that the optical properties of the system will be identical when rotating the whole structure by $\frac{2\pi}{n}$ radians.  But, as we analytically prove, the extinction, scattering and even absorption cross-sections are all identical for rotations of {\em any} angle.  Such structures can therefore be considered as being  polarisation-independent.

An interesting corollary to this result is that, while the extinction and scattering cross-sections are defined only in the far-field, the absorption can be calculated by two independent ways - as an energy balance between the far-field scattered and incident fields, and as integration of losses in the near-field. Both approaches produce the same result. In the near-field, the full profile of the electromagnetic field should be taken into account, while in the far-field only the leading order will survive. The polarisation-independent absorption is then quite counter-intuitive because the near-field profile of the electromagnetic field does depend on the incident polarisation yet the overall absorption cross-section does not.  It implies that the variation of the near-field with  incident polarisation does not affect the overall integral absorption cross-section. Thus, the near-field distribution still contains some symmetry properties of the entire structure.

In this paper we study optical properties of symmetric systems consisting of small ($R \ll \lambda$) spherical particles, which can be approximated as electric and magnetic dipoles using the coupled dipole approximation~\cite{Mulholland1994_Langmuir}. Given that any structure can ultimately be decomposed into such particles and subsequently a composite of electric and magnetic dipoles; the derived polarisation-independent response should be a typical feature of any system exhibiting {\em n}-fold rotational symmetry. As this is a purely geometric feature and shows no dependence on either the wavelength, the optical properties of the constituent materials or the resonances excited, we expect that it could be applied in various applications including sensing, imaging, photovoltaic devices and other biological and medical research.

\section{Theoretical model}

The scattering and absorption of light by a small spherical particle
$(R\ll\lambda)$ can be approximated by that of an electric and magnetic
dipole with electric and magnetic dipole moments, $\mathbf{p},\mathbf{\, m}$, given by
\begin{subequations}\label{eq:dipole moment definitions}
{\small \begin{align}
\mathbf{p} & =  \alpha_{E}\epsilon_{0}\mathbf{E} \\
\mathbf{m} & =  \alpha_{H}\mathbf{H}\;
\end{align}}%
\end{subequations}
where  $\mathbf{E},\,\mathbf{H}$ are the electric and magnetic
fields acting on the particle and $\alpha_{E},\,\alpha_{H}$ are the
scalar effective polarisabilities as defined below in terms of Mie Theory dipole scattering coefficients $a_{1},\: b_{1}$.

\begin{subequations}\label{eq:dipole moment definitions_2}
{\small \begin{align}
\alpha_{E} & =  6i\pi a_{1}/k^{3} \\
\alpha_{H} & =  6i\pi b_{1}/k^{3}
\end{align}}%
\end{subequations}

In this approximation the scattered fields from a single particle in free space are then described using dyadic Green's functions
\begin{subequations}\label{eq:scattered field single particle}
{\small \begin{align}
\mathbf{E^{s}}(\mathbf{r}) & =  { k^{2}\left[\frac{1}{\epsilon_{0}}\hat{G}^{0}(\mathbf{r},\mathbf{r}_{0})\mathbf{p}-\sqrt{\frac{\mu_{0}}{\epsilon_{0}}}\left(\nabla\times\hat{G}^{0}(\mathbf{r},\mathbf{r}_{0})\right)\mathbf{m}\right]} \\
\mathbf{H^{s}}(\mathbf{r}) & =  { k^{2}\left[\hat{G}^{0}(\mathbf{r},\mathbf{r}_{0})\mathbf{m}+\frac{1}{\sqrt{\epsilon_{0}\mu_{0}}}\left(\nabla\times\hat{G}^{0}(\mathbf{r},\mathbf{r}_{0})\right)\mathbf{p}\right]}
\end{align}}%
\end{subequations}
where free space dyadic Green's functions are defined as
\begin{subequations}\label{eq:Cij and fij definitions}
{\small \begin{align}
\hat{G}^{0}(\mathbf{r}_{i},\mathbf{r}_{j}) & =  \hat{G}_{ij}^{(3)}=a_{ij}\hat{I}^{(3)}+b_{ij}\mathbf{n}_{ji}\otimes\mathbf{n}_{ji} \\
\nabla\times\hat{G}^{0}(\mathbf{r}_{i},\mathbf{r}_{j}) & =  \hat{g}_{ij}^{(3)}=d_{ij}\left(\mathbf{n}_{ji}\otimes\mathbf{n}_{ji}-\hat{I}^{(3)}\right)^{\frac{1}{2}}
\end{align}}%
\end{subequations}
In these the  $\mathbf{n}_{ji}\times\hat{I}^{(3)}$ operator is expressed explicitly as
$\left(\mathbf{n}_{ji}\otimes\mathbf{n}_{ji}-\hat{I}^{(3)}\right)^{\frac{1}{2}}$  and we have defined scalars
\begin{subequations}\label{eq:CDE coefficients}
{\small \begin{align}
a_{ij} & =  { \frac{e^{ik r_{ji}}}{4\pi r_{ji}}\left(1+\frac{i}{k r_{ji}}-\frac{1}{k^{2}r_{ji}^{2}}\right)} \\
b_{ij} & =  { \frac{e^{ik r_{ji}}}{4\pi r_{ji}}\left(-1-\frac{3i}{k r_{ji}}+\frac{3}{k^{2}r_{ji}^{2}}\right)} \\
d_{ij} & =  { \frac{e^{ik r_{ji}}}{4\pi r_{ji}}\left(1+\frac{i}{k r_{ji}}\right)}
\end{align}}%
\end{subequations}
where $r_{ji}=|\mathbf{r}_{i}-\mathbf{r}_{j}|$ and $\mathbf{n}_{ji}=\frac{1}{r_{ji}}\left(\mathbf{r}_{i}-\mathbf{r}_{j}\right)$.

In an arbitrary system constructed from \emph{N} particles
the fields acting on the \emph{i}th particle will be the sum of both
the externally-applied incident fields and the scattered fields from
all the other particles. Therefore the expressions for the dipole moments of the \emph{i}th particle are
\begin{subequations}\label{eq:original dipole equations}
{\small \begin{align}
\mathbf{p}_{i} & =  { \alpha_{E,i}\epsilon_{0}\mathbf{E^{0}}_{i}+\alpha_{E,i}\epsilon_{0}k^{2}\left(\underset{i\neq j}{\overset{N}{\sum}}\frac{1}{\epsilon_{0}}\hat{G}_{ij}^{(3)}\mathbf{p}_{j}-\sqrt{\frac{\mu_{0}}{\epsilon_{0}}}\hat{g}_{ij}^{(3)}\mathbf{m}_{j}\right)} \\
\mathbf{m}_{i} & =  { \alpha_{H,i}\mathbf{H^{0}}_{i}+\alpha_{H,i}k^{2}\left(\underset{i\neq j}{\overset{N}{\sum}}\hat{G}_{ij}^{(3)}\mathbf{m}_{j}+\frac{1}{\sqrt{\epsilon_{0}\mu_{0}}}\hat{g}_{ij}^{(3)}\mathbf{p}_{j}\right)}
\end{align}}%
\end{subequations}
where $\mathbf{E^{0}}_{i}=\mathbf{E}^{0}(\mathbf{r}_{i})$, $\mathbf{H^{0}}_{i}=\mathbf{H^{0}}(\mathbf{r}_{i})$
are the externally-applied electric and magnetic fields at $\mathbf{r}_{i}$.

The dipole equations in Eq.~\ref{eq:original dipole equations} show
that each incident field vector can be equated to some linear combination
of the dipole moments and, as such, there will exist some $6N\times6N$  \emph{interaction matrix},
 $\hat{M}^{(6N)}$, to relate all the dipole moments to the incident field vectors.
To write this relation in the form of a matrix equation we define a state consisting of
all dipole moments, $\left|p\right\rangle $, and a state consisting
the incident fields acting on each dipole, $\left|e\right\rangle $.
\begin{subequations}\label{eq:state definitions}
{\small \begin{align}
\left|p\right\rangle  & =  {\scriptstyle \left(\begin{array}{cccccc}
\mathbf{p}_{1} & \cdots  \mathbf{p}_{N} & \sqrt{\epsilon_{0}\mu_{0}}\mathbf{m}_{1} & \cdots & \sqrt{\epsilon_{0}\mu_{0}}\mathbf{m}_{N}\end{array}\right)} \\
\left|e\right\rangle  & =  {\scriptstyle \left(\begin{array}{cccccc}
\mathbf{E^{0}}_{1} & \cdots  \mathbf{E^{0}}_{N} & \mathbf{\sqrt{\frac{\mu_{0}}{\epsilon_{0}}}}\mathbf{H^{0}}_{1} & \cdots  \mathbf{\mathbf{\sqrt{\frac{\mu_{0}}{\epsilon_{0}}}}}\mathbf{H^{0}}_{N}\end{array}\right).}
\end{align}}%
\end{subequations}
The matrix equation that relates the dipole moments to the incident field is then written as
{\small \begin{equation}
\epsilon_{0}\mathbf{\hat{\alpha}}^{(6N)}\left|e\right\rangle =\hat{M}^{(6N)}\left|p\right\rangle \label{eq:matrix CDE}
\end{equation}}%
where $\mathbf{\hat{\alpha}}^{(6N)}$ is a diagonal matrix containing
the electric and magnetic scalar polarisabilities, $\alpha_{E}$ and
$\alpha_{H}$, of each particle.

To generate a general expression for the interaction matrix, we define four smaller matrices;
two that couple electric or magnetic dipoles together, $\hat{M}_{ee}^{(3N)}$
and $\hat{M}_{hh}^{(3N)}$, and two that couple electric to magnetic
dipoles and vice versa, $\hat{M}_{eh}^{(3N)}$ and $\hat{M}_{he}^{(3N)}$.
These matrices are constructed from the dyadic Green's functions
as to match the concatenation of vectors in $\left|p\right\rangle $
and $\left|e\right\rangle $ [see Eq.~\ref{eq:state definitions}]
\begin{subequations} \label{eq:M construction start}
{\small \begin{align}
{ \hat{M}_{ee}^{(3N)}} & =  k^{2}\hat{\mathbf{\alpha}}_{E}{ \hat{M}_{G}^{(3N)}} \\
{ \hat{M}_{eh}^{(3N)}} & =  -k^{2}\hat{\mathbf{\alpha}}_{E}{ \hat{M}_{g}^{(3N)}} \\
{ \hat{M}_{he}^{(3N)}} & =  k^{2}\hat{\mathbf{\alpha}}_{H}{ \hat{M}_{G}^{(3N)}} \\
{ \hat{M}_{hh}^{(3N)}} & =  k^{2}\hat{\mathbf{\alpha}}_{H}{ \hat{M}_{g}^{(3N)}}
\end{align}}%
\end{subequations}
where $\hat{\mathbf{\alpha}}_{E,H}$ is a diagonal matrix containing
the electric or magnetic scalar polarisability of each particle and ${ \hat{M}_{G}^{(3N)},\, \hat{M}_{g}^{(3N)}}$ are defined as
\begin{subequations} \label{eq:Mg and MG definitions}
{\small \begin{align}
{ \hat{M}_{G}^{(3N)}} & =  \left(\begin{array}{cccc}
\hat{0}^{(3)} & -\hat{G}_{12}^{(3)} & \cdots & -\hat{G}_{1N}^{(3)}\\
-\hat{G}_{12}^{(3)} & \hat{0}^{(3)} &  & \vdots\\
\vdots &  & \ddots\\
-\hat{G}_{1N}^{(3)} & \cdots &  & \hat{0}^{(3)}
\end{array}\right)\\
{ \hat{M}_{g}^{(3N)}} & =  \left(\begin{array}{cccc}
\hat{0}^{(3)} & -\hat{g}_{12}^{(3)} & \cdots & -\hat{g}_{1N}^{(3)}\\
\hat{g}_{12}^{(3)} & \hat{0}^{(3)} &  & \vdots\\
\vdots &  & \ddots\\
\hat{g}_{1N}^{(3)} & \cdots &  & \hat{0}^{(3)}
\end{array}\right).
\end{align}}%
\end{subequations}

The complete interaction matrix as defined in Eq.~\ref{eq:matrix CDE} to relate incident field to dipole moments is then constructed as
{\small \begin{equation}
\hat{M}^{(6N)}=\hat{I}^{(6N)}+\left(\begin{array}{cc}
\hat{M}_{ee}^{(3N)} & \hat{M}_{eh}^{(3N)}\\
\hat{M}_{he}^{(3N)} & \hat{M}_{hh}^{(3N)}
\end{array}\right).\label{eq:M construction end}
\end{equation}}%
As such all further analysis will use the matrix equation, Eq.~\ref{eq:matrix CDE},  to study the optical responses of particle systems.

\section{Polarisation invariance and symmetry}

In this section we derive an expression for the commutation relation between the interaction matrix (see Eq.~\ref{eq:matrix CDE}) and a generic symmetry operation in order to implement group theory principles and restrict the shape of the interaction matrix based on the symmetry of a given system.  We then consider the case of general particle systems that have
a rotational symmetry described by an {\em n-fold axis}; an axis about which any number of $\frac{2\pi}{n}$ rotations will leave the system unchanged. This symmetry is also referred to as
 {\em cyclic symmetry} and the corresponding group of operations that represent this symmetry are denoted $C_{n}$ (the {\em cyclic group}). In doing this, we will show that the extinction, scattering and absorption cross-sections are all independent of the incident field polarisation in any system with cyclic symmetry.

In the coupled dipole equations (Eq.~\ref{eq:original dipole equations}), the only terms that contain information on the geometrical structure of a system
are the dyadic Green's functions, $\hat{G}_{ij}^{(3)}$ and $\hat{g}_{ij}^{(3)}$.  Moreover, it follows from the definitions in Eq.~\ref{eq:Cij and fij definitions}
that both $\hat{G}_{ij}^{(3)}$ and $\hat{g}_{ij}^{(3)}$ will transform
as a change of basis when any unitary operation, $\hat{U}^{(3)}$, is applied uniformly to a system's position
vectors
\begin{subequations} \label{eq:gij evolution}
{\small \begin{align}
\hat{G}_{ij}^{(3)} & \rightarrow  \hat{U}^{(3)}\hat{G}_{ij}^{(3)}\left(\hat{U}^{(3)}\right)^{\dagger} \\
\hat{g}_{ij}^{(3)} & \rightarrow  \hat{U}^{(3)}\hat{g}_{ij}^{(3)}\left(\hat{U}^{(3)}\right)^{\dagger}.
\end{align}}%
\end{subequations}
For $\hat{g}_{ij}^{(3)}$ to transform in this way,  we must also acknowledge that  it  can always be expanded as a Taylor series.  This is true because the magnitude of every component in $\left(\mathbf{n}_{ji}\otimes\mathbf{n}_{ji}-\hat{\mathbf{I}}\right)$ will be less than or equal to one and therefore a Taylor series for $\left(\mathbf{n}_{ji}\otimes\mathbf{n}_{ji}-\hat{\mathbf{I}}\right)^{\frac{1}{2}}$, and hence $\hat{g}_{ij}^{(3)}$,  will always converge.
So the overall interaction matrix must also transform in an analogous manner given its construction in terms of dyadic Green's functions (shown in Eq.~\ref{eq:M construction start}-\ref{eq:M construction end}) and because every $\hat{G}_{ij}^{(3)}$ and $\hat{g}_{ij}^{(3)}$ will transform uniformly according to Eq.~\ref{eq:gij evolution}
{\small \begin{equation}
\begin{array}{ccc}
\hat{M}^{(6N)} & \rightarrow & \hat{U}^{(6N)}\hat{M}^{(6N)}\left(\hat{U}^{(6N)}\right)^{\dagger}\end{array}
\end{equation}}%
where $\hat{U}^{(6N)}$ is defined such that it applies $\hat{U}^{(3)}$
to the position vector of every particle in $\hat{M}^{(6N)}$
{\small \begin{align}
\hat{U}^{(6N)} & =  \hat{U}^{(3)}\hat{I}^{(6N)}. \label{eq:aggregate 3x3 operator}
\end{align}}%
However, for the case of a unitary symmetry operation, $\hat{R}^{(3)}$, it also follows that each transformed position vector must be one of the existing position vectors. That is to say
{\small \begin{equation}
\hat{R}^{(3)}\mathbf{r}_{i}=\mathbf{r}_{j},\quad\mathrm{for\;each\;}i,\,j\in\{1,...N\}.
\end{equation}}%
Therefore we can express Eq.~\ref{eq:gij evolution} in the following manner
\begin{subequations}
{\small \begin{align}
\begin{array}{c}
\hat{R}^{(3)}\hat{G}_{ij}^{(3)}\left(\hat{R}^{(3)}\right)^{\dagger}\end{array} & =  \hat{G}_{kl}^{(3)} \\
\hat{R}^{(3)}\hat{g}_{ij}^{(3)}\left(\hat{R}^{(3)}\right)^{\dagger} & =  \hat{g}_{kl}^{(3)}.
\end{align}}%
\end{subequations}
Noticeably this means that $\hat{R}^{(3)}$ will not necessarily act as a symmetry operation on these dyadic Green's functions despite being a symmetry operation on the structure.
It does, however, show that there must exist a single permutation matrix, $\hat{\Pi}^{(3N)}$,
for each $\,\hat{R}^{(3)}$  such that
\begin{subequations} \label{eq:Mg MG commutation relation}
{\small \begin{align}
\begin{array}{c}
\hat{R}^{(3N)}\hat{M}_{G}^{(3N)}\left(\hat{R}^{(3N)}\right)^{\dagger}\end{array} & =  \left(\hat{\Pi}^{(3N)}\right)^{T}\hat{M}_{G}^{(3N)}\hat{\Pi}^{(3N)} \\
\hat{R}^{(3N)}\hat{M}_{g}^{(3N)}\left(\hat{R}^{(3N)}\right)^{\dagger} & = \left(\hat{\Pi}^{(3N)}\right)^{T}\hat{M}_{g}^{(3N)}\hat{\Pi}^{(3N)}
\end{align}}%
\end{subequations}
where $\,\hat{M}_{G}^{(3N)}$ and $\,\hat{M}_{g}^{(3N)}$ are the quadrants of $\,\hat{M}^{(6N)}$ as defined in  Eq.~\ref{eq:Mg and MG definitions} and $\,\hat{R}^{(3N)}$ is defined in a manner analogous to  Eq.~\ref{eq:aggregate 3x3 operator}.
It then follows that there will also exist a permutation matrix, $\hat{\Pi}^{(6N)}$, which is constructed from four identical $\,\hat{\Pi}^{(3N)}$ quadrants and satisfies
{\small \begin{equation}
\hat{R}_{n}^{(6N)}\hat{M}^{(6N)}\left(\hat{R}_{n}^{(6N)}\right)^{\dagger}=\left(\hat{\Pi}_{n}^{(6N)}\right)^{T}\hat{M}^{(6N)}\hat{\Pi}_{n}^{(6N)}.\label{eq:rotation to permutation}
\end{equation}}%
We can consider all non-zero components in these permutation matrices as
 $\hat{I}^{(3)}$ matrices and therefore they will always commute with the $\hat{R}$ matrices and also the $\hat{\alpha}^{(6N)}$ matrices defined for Eq.~\ref{eq:matrix CDE}.
It is then straightforward to rearrange Eq.~\ref{eq:Mg MG commutation relation}  and Eq.~\ref{eq:rotation to permutation} to show that
the general, symmetric commutation relation of the interaction matrix is
{\small \begin{align}
\hat{R}^{(6N)}\hat{\Pi}^{(6N)}\hat{M}^{(6N)} &=\hat{M}^{(6N)}\hat{R}^{(6N)}\hat{\Pi}^{(6N)}.\label{eq:R.Pi commutes with M}
\end{align}}%

In order to now consider the aggregate optical response of a symmetric system we refer to the general
expressions for the extinction, absorption and scattering cross-sections of any arbitrary system
with dipole moments $\left|p\right\rangle $ and incident field $\left|e\right\rangle $ (defined in Eq.~\ref{eq:state definitions})
\begin{subequations} \label{eq:original cross-section expressions}
{\small \begin{align}
\sigma_{e} & =  \frac{k}{\epsilon_{0}\left|\mathbf{E}^{0}\right|^{2}}\mathrm{Im}\left\{ \left\langle e|p\right\rangle \right\} \\
\sigma_{a} & =  { -\frac{k}{\left.\epsilon_{0}\right.^{2}\left|\mathbf{E}^{0}\right|^{2}}\left(\frac{k^{3}}{6\pi}\left\langle p|p\right\rangle +\mathrm{Im}\left\{ \left\langle p\right|{\hat{\alpha}}^{-1}\left|p\right\rangle \right\} \right)} \\
\sigma_{s} & =  \sigma_{e}-\sigma_{a}.
\end{align}}%
\end{subequations}
It is then desirable to re-express all three cross-sections in terms of the incident field state and three distinct matrices using Eq.~\ref{eq:matrix CDE}
\begin{subequations} \label{eq:sigma inner products}
{\small \begin{align}
\sigma_{e}  &= \mathrm{Im}\left\{
\left\langle e\right|\hat{M}_{1}'^{(6N)}\left|e\right\rangle \right\}\\
\sigma_{a}  &=
 \left\langle e\right|\hat{M}_{2}'^{(6N)}\left|e\right\rangle
+\mathrm{Im}\left\{ \left\langle e\right|\hat{M}_{3}'^{(6N)}\left|e\right\rangle\right\}\\
\sigma_{s}  &=  \sigma_{e}-\sigma_{a}
\end{align}}%
\end{subequations}
where we have defined the $\hat{M}'^{(6N)}$ matrices as
\begin{subequations}
{\small \begin{align}
\hat{M}_{1}'^{(6N)}&=\frac{k}{\left|\mathbf{E}^{0}\right|^{2}} (\hat{M}^{(6N)})^{-1}\hat{\alpha}^{(6N)}\\
\hat{M}_{2}'^{(6N)}&=-\frac{k^4 \epsilon_{0} }{6\pi\left|\mathbf{E}^{0}\right|^{2}} \left((\hat{M}^{(6N)})^{-1}\hat{\alpha}^{(6N)}\right)^{\dagger}(\hat{M}^{(6N)})^{-1}\hat{\alpha}^{(6N)}\\
\hat{M}_{3}'^{(6N)}&=-\frac{k\epsilon_{0}}{\left|\mathbf{E}^{0}\right|^{2}}\left((\hat{M}^{(6N)})^{-1}\hat{\alpha}^{(6N)}\right)^{\dagger}\left(\hat{M}^{(6N)} \hat{\alpha}^{(6N)}\right)^{-1}\hat{\alpha}^{(6N)}.
\end{align}}%
\end{subequations}
Given that $\,\hat{R}^{(6N)}\hat{\Pi}^{(6N)}$ will commute
with both the $\hat{M}^{(6N)}$ and $\hat{\alpha}^{(6N)}$ matrices,
 it therefore follows that $\,\hat{R}^{(6N)}\hat{\Pi}^{(6N)}$  must also commute with any of the
$\hat{M}'^{(6N)}$ matrices we have just defined. That is to say
all three inner products seen in Eq.~\ref{eq:sigma inner products} can ultimately be written in the one form
{\small \begin{align}
&\left\langle e\right|\hat{M}'^{(6N)}\left|e\right\rangle  \label{eq:M' commutivity} \\ 
\mathrm{where}\;\hat{R}^{(6N)}&\hat{\Pi}^{(6N)}\;\mathrm{commutes\;with}\;\hat{M}'^{(6N)}. \nonumber 
\end{align}}%
For this reason we can analyse the effect of the symmetric commutation relation (Eq.~\ref{eq:R.Pi commutes with M}) on all cross-sections at
once by considering an arbitrary $\hat{M}'^{(6N)}$ and evaluating Eq.~\ref{eq:M' commutivity}.
Moreover, to begin this analysis we separate the $\hat{M}'^{(6N)}$ matrix into the quadrants that
act between the electric and/or magnetic fields
{\small \begin{align}
\hat{M}'{}^{(6N)} & =\left(\begin{array}{cc}
\hat{M}'_{Q11}{}^{(3N)} & \hat{M}'_{Q12}{}^{(3N)}\\
\hat{M}'_{Q21}{}^{(3N)} & \hat{M}'_{Q22}{}^{(3N)}
\end{array}\right).\label{eq:quadrants of M'}
\end{align}}%
By doing this we can write the general inner product of Eq.~\ref{eq:M' commutivity} in terms
of the individual electric and magnetic field states
{\small \begin{align}
\left\langle e\right|\hat{M}'^{(6N)}\left|e\right\rangle  & =\left\langle e'\right|\hat{M}'_{Q11}{}^{(3N)}\left|e'\right\rangle +\left\langle e'\right|\hat{M}'_{Q12}{}^{(3N)}\left|h'\right\rangle   \nonumber
\\
 +&\left\langle h'\right|\hat{M}'_{Q21}{}^{(3N)}\left|e'\right\rangle
+\left\langle h'\right|\hat{M}'_{Q22}{}^{(3N)}\left|h'\right\rangle \label{eq:divideIntoComponents}
\\
\mathrm{where}  \left|e'\right\rangle &\equiv { \left(\begin{array}{c}
{ \mathbf{E^{0}}_{1}}\\
{ \vdots}\\
{ \mathbf{E^{0}}_{N}}
\end{array}\right)}\;\mathrm{and}\;\;\left|h'\right\rangle \equiv { \mathbf{{\textstyle \mathbf{\sqrt{\frac{\mu_{0}}{\epsilon_{0}}}}}}\left(\begin{array}{c}
{ \mathbf{H^{0}}_{1}}\\
{ \vdots}\\
{ \mathbf{H^{0}}_{N}}
\end{array}\right)}. \nonumber
\end{align}}%
 This can be simplified further if we only consider incident fields that are plane waves.  Moreover, we can define a new matrix, $\,\hat{A}^{(3N)}$, that will
introduce the appropriate phase differences to relate the field at each particle to a single field vector of the incident plane wave
 for the electric and magnetic components.  That is to say we define $\,\hat{A}^{(3N)}$ as
{\small \begin{align}
& \qquad \hat{A}^{(3N)} =\left(\begin{array}{cccc}
a_{11}\hat{I}^{(3)}\\
 & a_{22}\hat{I}^{(3)}\\
 &  & \ddots\\
 &  &  & a_{NN}\hat{I}^{(3)}
\end{array}\right) \label{eq:A(3N) construction}\\
&\mathrm{such\;that} \quad \left|e'\right\rangle  =\hat{A}^{(3N)}\left|e_{0}\right\rangle \quad \mathrm{and}\quad\left|h'\right\rangle  =\hat{A}^{(3N)}\left|h_{0}\right\rangle \nonumber \\
&\mathrm{where}\quad\left|e_{0}\right\rangle \equiv {\scriptstyle \left(\begin{array}{c}
{\textstyle \mathbf{E^{0}}}\\
{\textstyle \vdots}\\
{\textstyle \mathbf{E^{0}}}
\end{array}\right)}\;\mathrm{and}\;\;\left|h_{0}\right\rangle \equiv {\scriptstyle \mathbf{{\textstyle \mathbf{\sqrt{\frac{\mu_{0}}{\epsilon_{0}}}}}}\left(\begin{array}{c}
{\textstyle \mathbf{H^{0}}}\\
{\textstyle \vdots}\\
{\textstyle \mathbf{H^{0}}}
\end{array}\right)}. \nonumber
\end{align}}%
We are then able to rewrite Eq.~\ref{eq:divideIntoComponents}
in terms of a single incident field vector for both electric and magnetic
fields
{\small \begin{align}
\left\langle  e\right|  \hat{M}'^{(6N)}\left|e\right\rangle   =&\left\langle e_{0}\right|\underset{\equiv\hat{EE}{}^{(3N)}}{\underbrace{\left(\hat{A}^{(3N)}\right)^{\dagger}\hat{M}'_{Q11}{}^{(3N)}\hat{A}^{(3N)}}}\left|e_{0}\right\rangle  \nonumber \\ 
&+\left\langle e_{0}\right|\underset{\equiv\hat{EH}{}^{(3N)}}{\underbrace{\left(\hat{A}^{(3N)}\right)^{\dagger}\hat{M}'_{Q12}{}^{(3N)}\hat{A}^{(3N)}}}\left|h_{0}\right\rangle
\nonumber \\
&+\left\langle h_{0}\right|\underset{\equiv\hat{HE}{}^{(3N)}}{\underbrace{\left(\hat{A}^{(3N)}\right)^{\dagger}\hat{M}'_{Q21}{}^{(3N)}\hat{A}^{(3N)}}}\left|e_{0}\right\rangle \nonumber \\
 &+\left\langle e_{0}\right|\underset{\equiv\hat{HH}{}^{(3N)}}{\underbrace{\left(\hat{A}^{(3N)}\right)^{\dagger}\hat{M}'_{Q22}{}^{(3N)}\hat{A}^{(3N)}}}\left|h_{0}\right\rangle. \label{eq:definition of EE EH etc}
\end{align}}%
However, if we now express the four inner products in Eq.~\ref{eq:definition of EE EH etc} as sums, it becomes apparent that we can express the whole equation in terms of four $3\times3$ matrices
{\small \begin{align}
\left\langle e\right|&\hat{M}'^{(6N)}\left|e\right\rangle   
= \mathbf{E_{0}}^{\dagger}\underset{\equiv\hat{M}'_{ee}{}^{(3)}}{\underbrace{\underset{ij}{\sum}\hat{EE}_{ij}^{(3)}}}\mathbf{E_{0}} +\mathbf{H_{0}}^{\dagger}\underset{\equiv\hat{M}'_{he}{}^{(3)}}{\underbrace{\sqrt{\frac{\mu_0}{\epsilon_0}}\underset{ij}{\sum}\hat{HE}_{ij}^{(3)}}}\mathbf{E_{0}}
\nonumber \\
&\quad+ \mathbf{E_{0}}^{\dagger}\underset{\equiv\hat{M}'_{eh}{}^{(3)}}{\underbrace{\sqrt{\frac{\mu_0}{\epsilon_0}}\underset{ij}{\sum}\hat{EH}_{ij}^{(3)}}}\mathbf{H_{0}}
+\mathbf{H_{0}}^{\dagger}\underset{\equiv\hat{M}'_{hh}{}^{(3)}}{\underbrace{\frac{\mu_0}{\epsilon_0}\underset{ij}{\sum}\hat{HH}_{ij}^{(3)}}}\mathbf{H_{0}} \label{eq:M'ee M'eh etc definitions}
\end{align}}%
where we have denoted the $3\times3$ matrices that make up $\hat{EE}^{(3N)},$
$\hat{HE}^{(3N)},$ $\hat{EH}^{(3N)}$ and $\hat{HH}^{(3N)}$ according
to row and column indices
{\small \begin{align*}
\hat{EE}{}^{(3N)} & =\left(\begin{array}{cccc}
\hat{EE}_{11}^{(3)} & \hat{EE}_{12} ^{(3)}& \cdots\\
\hat{EE}_{21} ^{(3)}& \hat{EE}_{22}^{(3)}\\
\vdots &  & \ddots\\
 &  &  & \hat{EE}^{(3)}_{NN}
\end{array}\right).
\end{align*} }%
In summary, we have shown that any of the inner products from Eq.~\ref{eq:sigma inner products} can be written in the form
{\small \begin{align}
\left\langle e\right|\hat{M}'_{i}{}^{(6N)}\left|e\right\rangle  =&\mathbf{E_{0}}^{\dagger}\hat{M}'_{ee,i}{}^{(3)}\mathbf{E_{0}}+\mathbf{H_{0}}^{\dagger}
\hat{M}'_{he,i}{}^{(3)}\mathbf{E_{0}}
\nonumber \\
&+\mathbf{E_{0}}^{\dagger}\hat{M}'_{eh,i}{}^{(3)}\mathbf{H_{0}}+\mathbf{H_{0}}^{\dagger}\hat{M}'_{hh,i}{}^{(3)}\mathbf{H_{0}}. \label{eq:reduced arbitrary cross-section}
\end{align}}%

It is relatively straightforward to show that each $\,\hat{M}'_{ee}{}^{(3)}$, $\hat{M}'_{he}{}^{(3)}$, $\hat{M}'_{eh}{}^{(3)}$ and $\hat{M}'_{hh}{}^{(3)}\,$ will commute with the symmetry operators $\,\hat{R}^{(3)}$ given that each $\,\hat{M}'^{(6N)}$ commuted with $\,\hat{R}^{(6N)}\hat{\Pi}^{(6N)}$.
Specifically, because $\hat{\Pi}^{(6N)}$ is constructed from four identical quadrants (see Eq.~\ref{eq:Mg MG commutation relation}), it follows that each of the quadrants of $\,\hat{M}'^{(6N)}$ must necessarily commute with  $\,\hat{R}^{(3N)}\hat{\Pi}^{(3N)}$.  In other words
{\small \begin{align}
\hat{R}^{(3N)}\hat{\Pi}^{(3N)}\hat{M}'_{Qij}{}^{(3N)}  &=\hat{M}'_{Qij}{}^{(3N)}\hat{R}^{(3N)}\hat{\Pi}^{(3N)}.
\end{align}}%
The transformation of each $\hat{M}'_{Qij}{}^{(3N)}$ with the $\hat{A}^{(3N)}$ matrix in Eq.~\ref{eq:definition of EE EH etc} does not effect the existing commutation relation given $\hat{A}^{(3N)}$ is diagonal and constructed of multiples of the $3\times3$ identity matrix and therefore commutes directly with both $\hat{R}^{(3N)}$ and $\hat{\Pi}^{(3N)}$.
The following sum over all $3\times3$ matrix components of $\hat{EE}^{(3N)}$, $\hat{HE}^{(3N)},$ $\hat{EH}^{(3N)}$ and $\hat{HH}^{(3N)}$ in Eq.~\ref{eq:M'ee M'eh etc definitions} will then  absorb and ignore the permutation matrix leaving the commutation relation solely in terms of $\hat{R}^{(3)}$
{\small \begin{align*}
\hat{R}^{(3N)}\hat{\Pi}^{(3N)}\hat{EE}^{(3N)}  =\hat{EE}^{(3N)}\hat{R}^{(3N)}\hat{\Pi}^{(3N)}
  \\ \overset{\mathrm{sum\;}}{\Longrightarrow} \hat{R}^{(3)}\left(\underset{ij}{\sum}\hat{EE}_{ij}\right)  =\left(\underset{ij}{\sum}\hat{EE}_{ij}\right)\hat{R}^{(3)}.
\end{align*} }%
As such the $\,\hat{M}'_{ee}{}^{(3)}$, $\hat{M}'_{he}{}^{(3)}$, $\hat{M}'_{eh}{}^{(3)}$ and $\hat{M}'_{hh}{}^{(3)}$ matrices will all commute with the symmetry operators $\hat{R}^{(3)}$.  From this point onward we will only be dealing with these $3\times3$ matrices, so can neglect the indices and other notation to write this commutation relation as simply
{\small \begin{align}
\hat{R}\hat{M} & =\hat{M} \hat{R}. \label{eq:the commutation relation}
\end{align}}%
The commutation relation in Eq.~\ref{eq:the commutation relation} can be used to deduce
constraints on the given matrix, $\hat{M}$, depending on which symmetry group $\hat{R}$
corresponds to.  In the following argument we will consider only the constraints arising from
the $C_{n}$ group, however there will be analogous procedures for the other symmetry groups.
In any case, the elements of the $C_{n}$ group can be expressed as
 {\small \begin{align}
\left\{ \hat{I},\begin{array}{cccc}
\hat{C}_{n}, & \hat{C}_{n}^{2}, & \cdots & \hat{C}_{n}^{n-1}\end{array}\right\}
\end{align}}%
where $\hat{C}_{n}$ is a rotation about the symmetry axis of $\frac{2\pi}{n}$.

For practicality this group can also be represented in a matrix form
with a Cartesian basis in which the $\boldsymbol{z}$-basis vector is parallel to the
symmetry axis.  The general form of a matrix element in such a representation
is expressed in terms of the irreducible representations
of the $C_{n}$ group as
{\small \begin{align}
\hat{R} & =\left(\begin{array}{cc}
 \hat{U}^{\dagger}\left(\begin{array}{cc}
R_{11} & 0\\
0 & R_{22}
\end{array}\right)\hat{U}\;  \hfill  \vline   \\  \hline
\hfill \vline  &      R_{33}
\end{array} \right).\label{eq:symmetry operator}
\end{align}}%
where  $R_{11}$ and $R_{22}$ are the first conjugate pair of one-dimensional (degenerate) irreducible representations ($E_{1}$), $R_{33}$ is the symmetric 1-dimensional irreducible representation ($A_{1}$) and $\hat{U}$ is a unitary matrix used to describe the appropriate similarity transform for a Cartesian basis.\cite{McWeeny1963,Cotton1990}
\begin{subequations}\label{eq:IR definitions}
{\small \begin{align}
&R_{11} = \left\{ \begin{array}{ccccc}
1,&\omega, & \omega^{2}, & \cdots & \omega^{n-1}\end{array}\right\} ,\,\,\omega=e^{i\frac{2\pi}{n}}\\
&R_{22}= R_{11}^{*}\\
&R_{33}= \left\{ \begin{array}{ccccc}
1,&1, & 1, & \cdots & 1\end{array}\right\} \\
&\hat{U}= \frac{1}{\sqrt{2}}\left(\begin{array}{cc}
1 & i\\
1 & -i
\end{array}\right)
\end{align}}%
\end{subequations}
It is important to acknowledge that this decomposition of $\hat{R}$ in terms of distinct irreducible representations requires that $n\geq3$ because there are precisely $n$, 1-dimensional irreducible representations in the $C_{n}$ group.  That is to say the $C_2$ symmetry group only has two such irreducible representations and subsequently a matrix representation of its operators can't be expressed in the manner of Eq.~\ref{eq:symmetry operator}.
To continue with the $C_n$, $n\geq3$ case; we now divide the $\hat{M}$ matrix in a manner corresponding
to that done for $\hat{R}$ in Eq.~\ref{eq:symmetry operator}
{\small \begin{align}
\hat{M} & =\left(\begin{array}{cc}
\begin{array}{ccc} & & \\ & \hat{M}_{11} &\\ & & \end{array}   \hfill \vline & \ \hat{M}_{12} \\ \hline
\begin{array}{ccc}  & \hat{M}_{21} & \end{array}  \hfill \vline &   M_{22}
\end{array}\right).
\end{align}}%
As such we can then write the commutation relation
in Eq.~\ref{eq:the commutation relation}  as the following four equations
\begin{subequations}
{\small \begin{align}
R_{33}M_{22} & =M_{22}R_{33}\label{eq:obvious}\\
\left(\begin{array}{cc}
R_{11} & 0\\
0 & R_{22}
\end{array}\right)\left(\hat{U}\hat{M}_{12}\right) & =\left(\hat{U}\hat{M}_{12}\right)R_{33}\label{eq:zero1}\\
R_{33}\left(\hat{M}_{21}\hat{U}^{\dagger}\right) & =\left(\hat{M}_{21}\hat{U}^{\dagger}\right)\left(\begin{array}{cc}
R_{11} & 0\\
0 & R_{22}
\end{array}\right)\label{eq:zero2}\\
\hat{U}^{\dagger}\left(\begin{array}{cc}
R_{11} & 0\\
0 & R_{22}
\end{array}\right)\hat{U}\hat{M}_{11} & =\hat{M}_{11}\hat{U}^{\dagger}\left(\begin{array}{cc}
R_{11} & 0\\
0 & R_{22}
\end{array}\right)\hat{U}.\label{eq:notObvious}
\end{align}}%
\end{subequations}
Eq.~\ref{eq:obvious} is trivial as $M_{22}$ and $R_{33}$ are both
scalars, hence there are no restrictions on $M_{22}$ and we can just consider
it as some scalar $B\in\mathbb{C}$. Eq.~\ref{eq:zero1} and Eq.~\ref{eq:zero2}
are not so trivial and describe four scalar relationships between irreducible representations of the form
{\small \begin{align}
R_{ii}a & =aR_{jj},\qquad\mathrm{where\;}i\neq j,\; a\in\mathbb{C}.\label{eq:scalar relationship}
\end{align}}%
Noticeably, the only way non-trivial solutions to this sort of equation could exist is if the distinct
irreducible representations were equal to each other. However it is obvious that
distinct irreducible representations cannot be equal to each other
by their definition (e.g. see Eq.~\ref{eq:IR definitions}) and hence
the only valid solution is that both sides of Eq.~\ref{eq:zero1}
and Eq.\ref{eq:zero2} are equal to zero. Then, given
$\hat{U}$ is invertible, we can conclude that both $\hat{M}_{21}$
and $\hat{M}_{12}$ must therefore consist of only zeros.
The final equation from the commutation relation is then Eq.~\ref{eq:notObvious}, which
we can similarly rearrange to relate the $R_{11}$ and $R_{22}$ irreducible
representations to each other
{\small \begin{align}
\left(\begin{array}{cc}
R_{11} & 0\\
0 & R_{22}
\end{array}\right)\underset{\equiv\hat{M}'}{\underbrace{\hat{U}\hat{M}_{11}\hat{U}^{\dagger}}} & =\underset{\equiv\hat{M}'}{\underbrace{\hat{U}\hat{M}_{11}\hat{U}^{\dagger}}}\left(\begin{array}{cc}
R_{11} & 0\\
0 & R_{22}
\end{array}\right).\label{eq:definition of M'}
\end{align}}%
We can see that the off-diagonal terms of $\hat{M}'$ must be zero
as they produce a scalar relationship in the form of Eq.~\ref{eq:scalar relationship}
between $R_{11}$ and $R_{22}$. As it happens, this is the only constraint we can
apply and subsequently $\hat{M}'$ will be of the form
{\small \begin{align}
\hat{M}' & =\left(\begin{array}{cc}
a & 0\\
0 & b
\end{array}\right)\qquad\mathrm{where}\; a,\, b\in\mathbb{C}.
\end{align}}%
We can then get the corresponding $\hat{M}_{11}$ directly from the
definition of $\hat{M}'$ in Eq.~\ref{eq:definition of M'}
\noindent \begin{center}
{\small \begin{align}
\hat{M}_{11} & =\hat{U}^{\dagger}\hat{M}'\hat{U}=\frac{1}{2}\left(\begin{array}{cc}
a+b & (a-b)i\\
(b-a)i & a+b
\end{array}\right) \nonumber\\ & \equiv\left(\begin{array}{cc}
A & C\\
-C & A
\end{array}\right)\qquad\mathrm{where}\; A,\, C\in\mathbb{C}.
\end{align}}%
 \par\end{center}
In conclusion each matrix from Eq.~\ref{eq:reduced arbitrary cross-section} in a system with $C_{n}$
symmetry will be of the form
{\small \begin{align}
\hat{M} & =\left(\begin{array}{ccc}
A & C & 0\\
-C & A & 0\\
0 & 0 & B
\end{array}\right)\qquad\mathrm{where}\;\;A,\, B,\, C\,\in\mathbb{C}.  \label{eq:shape of M}
\end{align}}%
However, if the incident field is propagating in the direction of the symmetry axis ($E^{0}_{z},\, H^{0}_{z}=0$) then we can relate  $\mathbf{H_{0}}$ to $\mathbf{E_{0}}$ as
{\small \begin{align}
\mathbf{H^{0}}= \sqrt{\frac{\epsilon_0}{\mu_0}} \left(\begin{array}{ccc}
0 & -1 & 0\\
1 & 0 & 0\\
0 & 0 & 0\end{array}\right)\mathbf{E^{0}}. \label{eq:E0 H0 relation}
\end{align}}%
The combination of electric and magnetic inner products in Eq.~\ref{eq:reduced arbitrary cross-section} can then be reduced to just a single, purely-electric, inner product
{\small \begin{align}
&\left\langle e\right|\hat{M}'^{(6N)}\left|e\right\rangle
 \nonumber\\
&=\mathbf{E_{0}}^{\dagger}\hat{M}'_{ee}\mathbf{E_{0}}+\mathbf{H_{0}}^{\dagger}
\hat{M}'_{he}\mathbf{E_{0}}+\mathbf{E_{0}}^{\dagger}\hat{M}'_{eh}\mathbf{H_{0}} 
 +\mathbf{H_{0}}^{\dagger}\hat{M}'_{hh}\mathbf{H_{0}} 
\nonumber\\
&= \mathbf{E_{0}}^{\dagger} \underset{\mathrm{new\;3\times3\;matrix}}{\underbrace{\left(\hat{M}'_{ee} + \sqrt{\frac{\epsilon_0}{\mu_0}}\left(\hat{M}''_{eh} + \hat{M}''_{he} \right)+ {\frac{\epsilon_0}{\mu_0}} \hat{M}'_{hh} \right)}}\mathbf{E_{0}}.
\end{align}}%
Here the double-prime indicates that the matrix has been multiplied by the matrix seen in Eq.~\ref{eq:E0 H0 relation}, which noticeably commutes with- and does not change the general form of any matrix defined as per Eq.~\ref{eq:shape of M}.  Therefore, with no $z$-component of the incident field and a single matrix in the form of Eq.~\ref{eq:shape of M}, any of the inner products used to define the cross-sections in Eq.~\ref{eq:sigma inner products} can be written as
{\small \begin{align}
&\left\langle e\right|\hat{M}'_{i}{}^{(6N)}\left|e\right\rangle 
\nonumber \\
&= 
\left(\begin{array}{cc}
{E_x}^* & {E_y}^*
 \end{array} \right)
\left(\begin{array}{cc}
A_i & C_i\\
-C_i & A_i 
\end{array}\right) 
\left(\begin{array}{c}
E_x\\
E_y \end{array}\right)
\nonumber \\
&= A_i \left(\left|E_{x}\right|^{2}+\left|E_{y}\right|^{2}\right) + C_i \left({E_x}^{*} E_{y} -  E_{x} {E_y}^*\right)\label{eq:result}
\\
&\qquad\qquad\mathrm{where}\;\;A_i,\, C_i\,\in\mathbb{C}. \nonumber
\end{align}}%
In this paper we consider only linearly-polarised light and so the $E_x$ and $E_y$ components are in phase.  As such, the term proportional to $C_i$ in Eq.~\ref{eq:result} can be removed and we are left with
{\small \begin{align}
\left\langle e\right|\hat{M}'_{i}{}^{(6N)}\left|e\right\rangle &= A_i \left(\left|E_{x}\right|^{2}+\left|E_{y}\right|^{2}\right) . \label{eq:really reduced inner product}
\end{align}}%
The cases of other polarisations will be addressed in another paper.  However, for linearly-polarised light, the expressions for the cross-sections in Eq.~\ref{eq:sigma inner products} can be simplified using Eq.~\ref{eq:really reduced inner product} to become
\begin{subequations}
{\small \begin{align}
\sigma_{e} &=  \mathrm{Im}\left\{A_{1}\right\}\left(\left|E_{x}\right|^{2}+\left|E_{y}\right|^{2}\right) \\
\sigma_{a} &= \left( A_{2}+\mathrm{Im}\left\{ A_{3}\right\}\right) \left(\left|E_{x}\right|^{2}+\left|E_{y}\right|^{2}\right) \\
\sigma_{s} & =  \sigma_{e} - \sigma_{a} = \left(\mathrm{Im}\left\{A_{1}-A_{3}\right\} - A_{2 }\right)\left(\left|E_{x}\right|^{2}+\left|E_{y}\right|^{2}\right).
\end{align}}%
\end{subequations}
Noticeably this shows that all the cross-sections are independent of polarisation.  Specifically, we have shown that any system with at least 3-fold cyclic symmetry ($C_3$) will have polarisation-independent extinction, scattering and absorption cross-sections for linearly-polarised plane waves traveling parallel to the symmetry axis.

While this is the main result of the paper, it is also worth noting that the derivations of Eq.~\ref{eq:reduced arbitrary cross-section} and Eq.~\ref{eq:the commutation relation} are for arbitrary symmetry operations and therefore provide a foundation on which to evaluate cross-sections for operations ($\hat{R}$)  corresponding to different symmetries.  In this way it is possible to, for instance, consider the tetrahedral ($T$) or pyramidal ($C_{nv}$) symmetry groups, both of which can be handled by directly applying Schur's Lemma\cite{McWeeny1963} to show isotropic and axial polarisation-independent cross-sections.

\section{Examples of light scattering}

\begin{figure}[t]
\centering
\centerline{\includegraphics[width=\columnwidth]{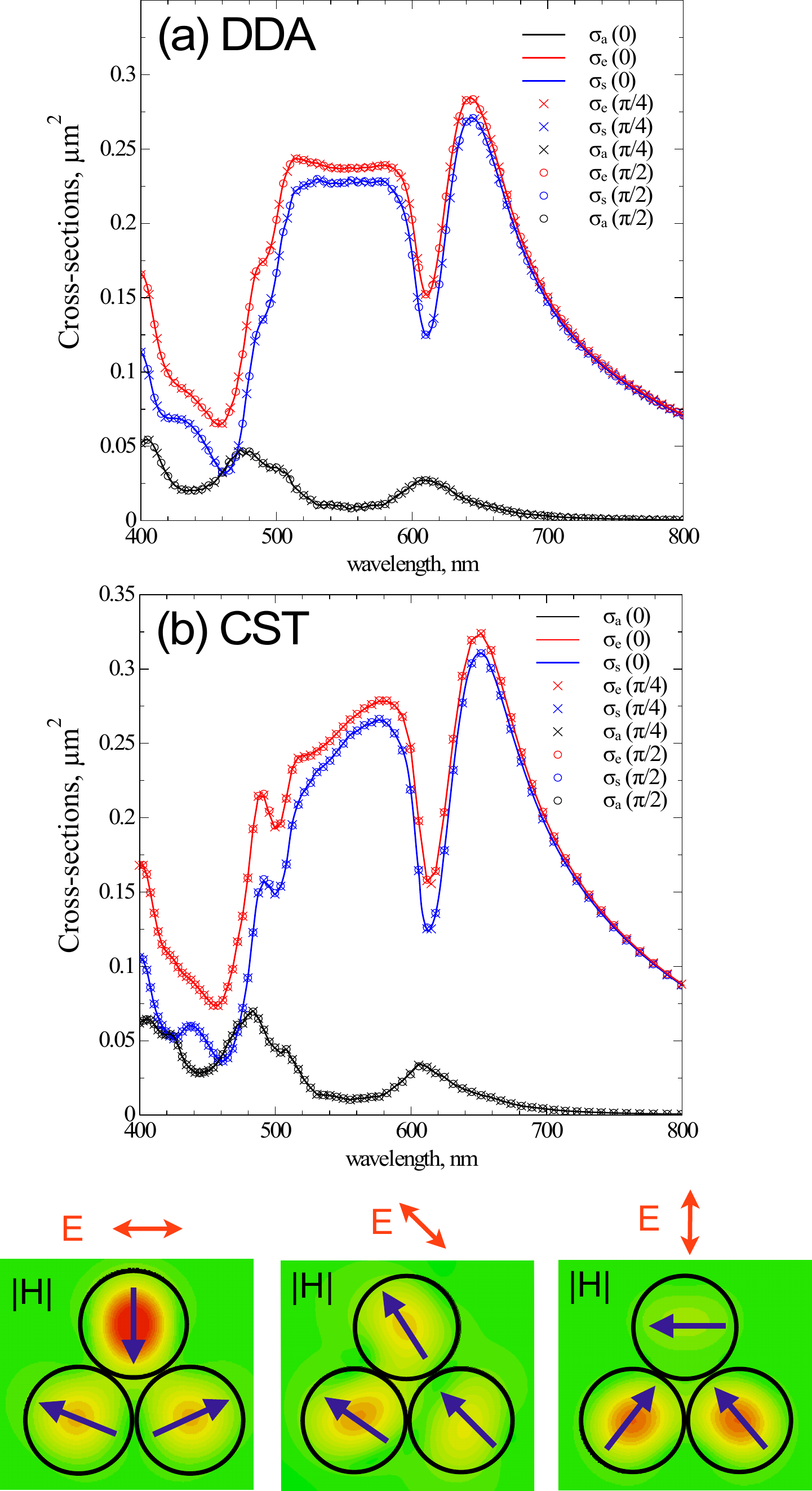}}
\caption{Extinction, scattering and absorption cross-sections for three different polarisations of a trimer structure made of touching silicon nanopartices of radii $R=75$nm calculated by using (a) the discrete dipole approximation (DDA) method and (b) direct numerical simulations with CST Microwave Studio. Bottom panels show magnetic field distribution at the, $\lambda=612$nm,  Fano resonance~\cite{Miroshnichenko:2012} for  all three polarisations together with the induced magnetic dipole moments.}
\label{fig:fig2}
\end{figure}

\begin{figure}[t]
\centering
\centerline{\includegraphics[width=\columnwidth]{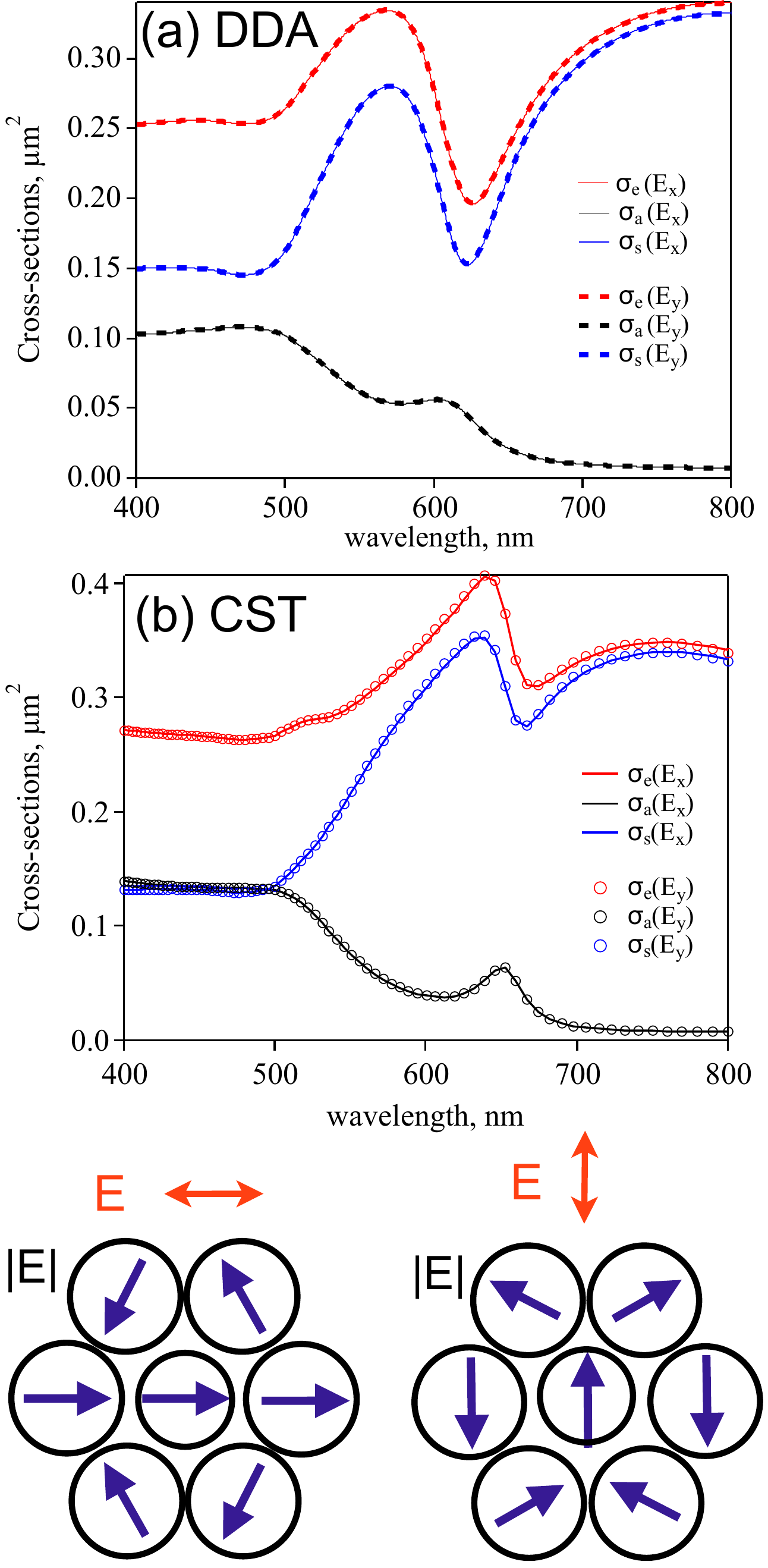}}
\caption{Extinction, scattering and absorption cross-sections for two orthogonal polarisations of a heptamer structure made of gold nanopartices of central $R_c=65$nm and outer $R_o=75$nm radii separated by $d=10$nm calculated by using (a) the discrete dipole approximation (DDA) method and (b) direct numerical simulations with CST Microwave Studio. Bottom panels show electric field distribution at the, $\lambda=600$nm, Fano resonance~\cite{Hentsche2010_NL} for two polarisations together with induced electric dipole moments.}
\label{fig:fig3}
\end{figure}

\begin{figure}[t]
\centering
\centerline{\includegraphics[width=\columnwidth]{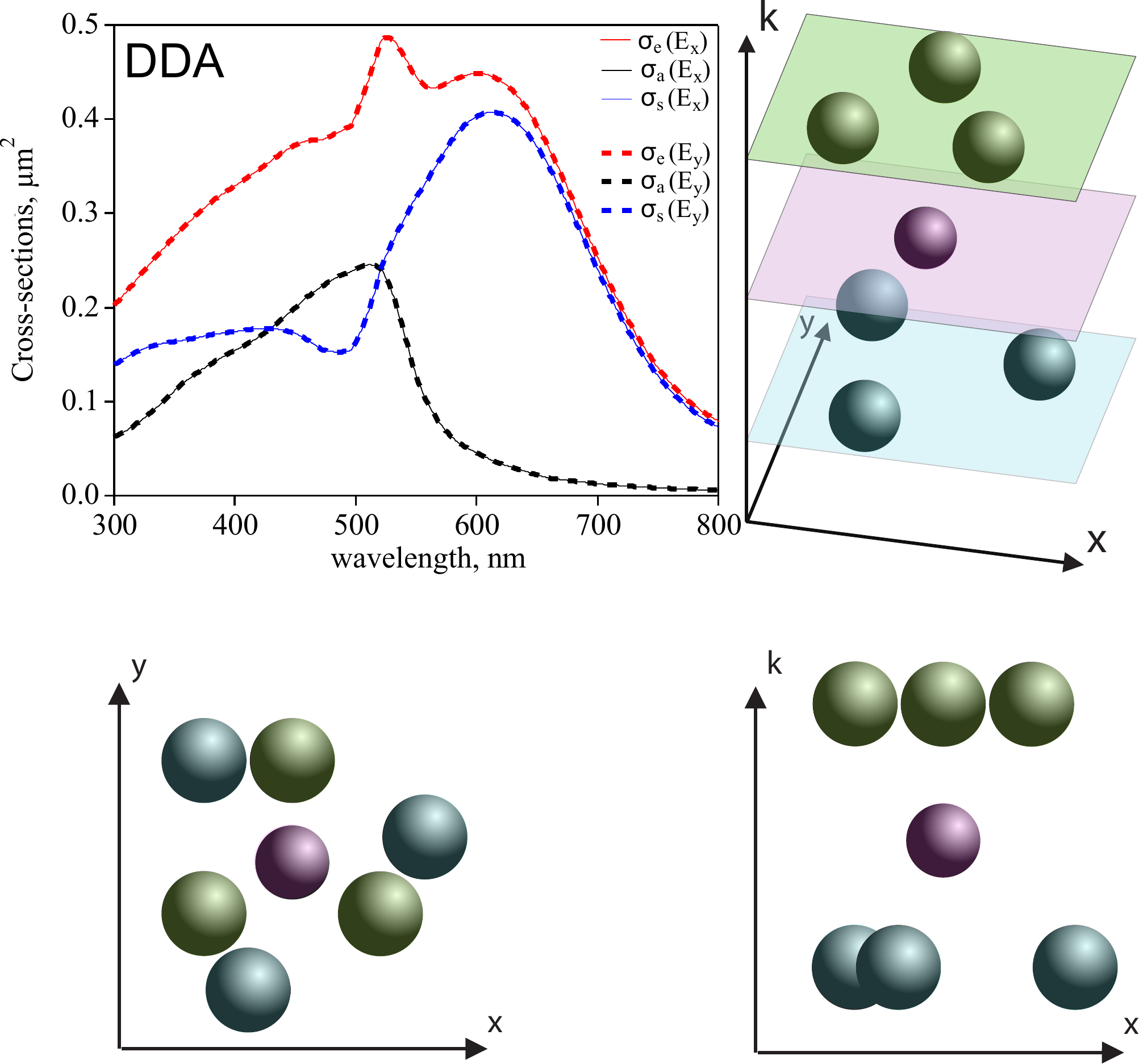}}
\caption{Extinction, scattering and absorption cross-sections for a structure with 3-fold symmetry consisting of two equilateral trimers (marked by different colors) that are separated by a central particle and evenly spaced along the propagation axis,  calculated by using the discrete dipole approximation (DDA). The trimers are additionally twisted with respect to each other in order to make the structure chiral and therefore satisfy only the minimum derived symmetry requirement needed to make the structure polarisation-independent.}
\label{fig:fig4}
\end{figure}

 To demonstrate the validity of our approach we employed two methods to study the light scattering by some oligomer structures with {\em n}-fold symmetries [see Figs. {\ref{fig:fig2}-\ref{fig:fig4}}]. Firstly we used CST Microwave Studio to calculate the exact total extinction, scattering and absorption cross-sections as well as the near-field profiles of the corresponding structures at resonance. And, secondly, we employed the dipole approximation and dyadic Green's function method to obtain the cross-sections and the distribution of optically-induced electric and magnetic dipoles in the individual nanoparticles~\cite{Mulholland1994_Langmuir}. 
Figure~\ref{fig:fig2} shows the extinction, scattering and absorption cross-sections of a trimer consisting of three silicon nanospheres with 3-fold symmetry. Figure~\ref{fig:fig3} shows the extinction, scattering and absorption cross-sections for a  heptamer consisting of seven gold nanospheres with 6-fold symmetry. All particles of these oligomer-like structures are in the same transverse plane, so the excitation field is identical for all particles. We can also lower the symmetry of a structure by shifting some particles along the propagation axis. For example, in Fig.~\ref{fig:fig4} we present a structure with 3-fold symmetry which was derived from a gold heptamer structure shown in Fig.~\ref{fig:fig3}. To construct the new structure we shifted two equilateral trimers of the outer ring in opposite directions from the central particle evenly spaced along the propagation axis, and then twisted each with respect to the other. The final structure is then chiral with 3-fold symmetry. Thus, according to our theoretical prediction, we expect that it should be polarisation-invariant. We used Palik's data for permittivity of the various materials~\cite{Palik}.
All the presented results support our derivation that these structures will exhibit polarisation-independent optical properties for {\em any} incident polarisation angle, which does not necessarily coincide with the rotational symmetry of the structures. These figures also show that the total absorption is polarisation-independent even though the near-field distribution varies with the incident polarisation. It allows us to conclude that, although all structures exhibit some degree of geometrical anisotropy, their optical response is {\em isotropic}. And the only requirement that we impose is that the structure supports {\em n}-fold symmetry with $n\ge3$. For completeness we also acknowledge that structures with only $C_2$ symmetry are known to be polarisation-dependent~\cite{Chuntonov2011_NL} and therefore can conclude that $n\ge3$ is a requirement for the {\em n}-fold symmetry.

\begin{figure}[t]
\centering
\centerline{\includegraphics[width=0.8\columnwidth]{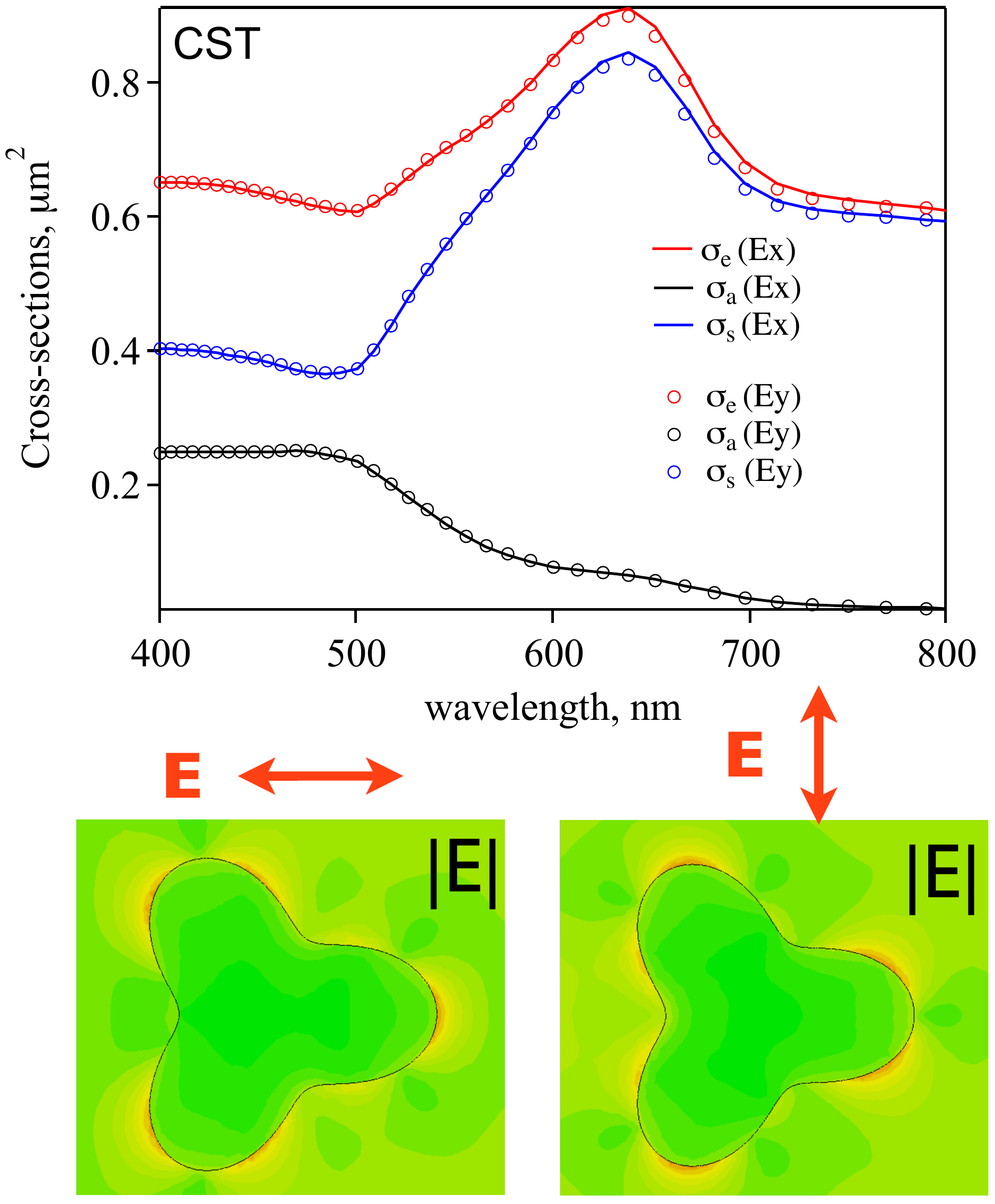}}
\caption{ Extinction, scattering and absorption cross-sections for two orthogonal polarisations of a continuous structure with 3-fold symmetry, calculated by using direct numerical simulations with CST Microwave Studio.}
\label{fig:fig5}
\end{figure}

Finally, based on the coupled dipole approximation method,~\cite{Draine:JOSAA:1994} the optical response of structures with {\em arbitrary} geometries and complex refractive index can be approximated  by an ensemble of  discrete dipoles. Thus, our results can be easily generalised  to {\em any} structure with {\em n}-fold symmetry. Figure~\ref{fig:fig5} shows the results of direct numerical simulations of a continuous structure with 3-fold symmetry, which, for simplicity, is modeled as $\rho(\theta)=R[1+\cos(1.5\theta)]$ with $R=200$~nm (on the transverse plane) and $h=100$~nm (along longitudinal direction) and is made of gold. It still exhibits polarisation-independent optical response, in full agreement with our approach above. This proves that our statement is quite universal and can be applied to any system.

\section{Conclusions}

 We have studied the optical response of nanoparticle structures with an {\em n}-fold ($n\geq3$) rotational symmetry excited by an incident plane wave propagating parallel to the symmetry axis. We have demonstrated  that polarisation-independent responses (in terms of the cross-section of scattering, absorption and extinction) can come solely from the overall rotational symmetry of a structure without any condition placed on other elements of a given system. We have presented specific examples which support our general theory.  Such robust polarisation-independent features are expected to play an important role in various applications including nanoantennas, sensing, imaging, solar cells, and other applications in chemistry, biology, and medicine.

\section{Acknowledgements}

The authors thank Anton Desyatnikov, Andrey Sukhorukov, Dragomir Neshev and Alexander Poddubny for useful discussions, and also acknowledge a partial support of the Future Fellowship program of the Australian Research Council (Project No. FT110100037).

\footnotesize{
\bibliographystyle{rsc}
\bibliography{References_scattering}
}

\end{document}